\documentclass[12pt,preprint]{aastex}

\usepackage{graphicx}
\usepackage{amssymb}
\usepackage{epstopdf}

\begin{document}

\title{Numerical Evidence for Dark Star Formation: A Comment on ``Weakly Interacting Massive Particle Dark Matter and First Stars: Suppression of Fragmentation in Primordial Star Formation'' by Smith et al.\ 2012, ApJ 761, 154}
\author{
Paolo Gondolo\altaffilmark{1},
Katherine Freese\altaffilmark{2},
Douglas Spolyar\altaffilmark{3},
and Peter Bodenheimer\altaffilmark{4}
}
\email{paolo.gondolo@utah.edu}
\email{ktfreese@umich.edu}
\email{dspolyar@iap.fr}
\email{peter@ucolick.org}

\altaffiltext{1}{Dept.\ of Physics and Astronomy, University of Utah, Salt Lake City, UT 84112}
\altaffiltext{2}{Michigan Center for Theoretical Physics, Physics Dept.,
Univ.\ of Michigan, Ann Arbor, MI 48109}
\altaffiltext{3}{Institut d'Astrophysique, Paris, France}
\altaffiltext{4}{UCO/Lick Observatory and Dept. of Astronomy and Astrophysics,
 University of California, Santa Cruz, CA 95064}

\begin{abstract}
\noindent
This comment is intended to show that simulations of Smith et al.\ support the Dark Star (DS) scenario and even remove some potential obstacles.  
Our previous work illustrated that the initial hydrogen densities of the first equilibrium dark stars that form are high, roughly $n_H \sim 10^{17}$ cm$^{-3}$ 
for the case of 100 GeV WIMPs, with a stellar radius of $\sim$ 2--3 AU.  Subsequent authors have somehow missed the fact that equilibrium DS have the high densities they do. Smith et al.\ have produced three-dimensional simulations to follow the effect of dark matter annihilation on the contraction of a protostellar gas cloud en route to forming the first stars.  They show results at $n_H \sim 5\times10^{14}~{\rm cm}^{-3}$, a density slightly higher than the value at which annihilation heating prevails over cooling.  However, they are apparently unable to reach the $\sim 10^{17}~{\rm cm}^{-3}$ density of our hydrostatic dark star solutions.   We  are in complete agreement with their physical result that the gas keeps collapsing to  densities higher than $n_H=5 \times 10^{14}$ cm$^{-3}$, as it must before equilibrium DS can form.   However we are in disagreement with some of the words in their paper which imply that dark stars never come to exist.  It seems to us that Smith et al.'s work supports the dark star scenario.  They use the sink particle approach to treat the gas that collapses to scales smaller than their resolution limit.  We argue that their sink is effectively a dark star (or contains one).  An accretion disk forms as more mass falls onto the sink, and the dark star grows.  The simulations of Smith et al.\ not only confirm our predictions about dark stars in the range where the simulations apply, but also solve a potential obstruction to dark star formation by showing that dark matter annihilation prevents the fragmentation of the collapsing gas.  Whereas one might worry that fragmentation might perturb the dark matter away from the dark star and remove its power source, instead Smith et al.\ show that further sinks, if any, form only far enough away as to leave the dark star undisturbed in the comfort of its dark matter surroundings.  
\end{abstract}
\keywords{Dark Matter, Star Formation, Accretion}

The idea that the formation history of the first stars in the universe may be modified by the energy of annihilating dark matter was first proposed in Spolyar, Freese \& Gondolo (2008), hereafter Paper I. In current theories of primordial star formation, the first stars (Population III stars) form in the central regions of dark matter halos by the 
gravitational contraction and energy dissipation of the baryonic gas  (for reviews, see, e.g., Barkana \& Loeb 2001, Yoshida et al.\ 2003, Bromm \& Larson 2004, and 
Ripamonti \& Abel 2004). Early theoretical calculations  indicated that the gas cools and collapses via H$_2$ cooling (Peebles \& Dicke 1968, Matsuda et al. 1971, 
Hollenbach \& McKee 1979) into a single small protostar at the center of the halo (Omukai \& Nishi 1998; Abel et al.\ 2000, 2002; Yoshida et al.\ 2003; Yoshida et al.\ 2008).  
Later simulations found that the gas may fragment into a close binary or multiple stellar system (e.g., Clark et al.\ 2008, 2011a, 2011b; Turk et al.\ 2009; Stacy et al.\ 2010; 
Greif et al.\ 2011, 2012). The simulations of Smith et al.\ (2012; hereafter Smith et al.) show instead that dark matter annihilation suppresses fragmentation and a central 
protostar is favored, either single or undisturbed by other fragments for very long times.

In Paper I, we took a single protostar at the center of the contracting gas and we modeled the concomitant contraction of the dark matter, which is dragged into the 
gravitational potential created by the gas. We realized that the dark matter may become so dense that the heat produced by the annihilation of dark matter particles inside 
the forming protostar may overtake the cooling of the contracting gas. The final products of the annihilation are predominantly electron/positron pairs and photons 
(neutrinos would escape the star), all of which get stuck in the gas (we demanded 80 radiation lengths), thermalize with the gas, and heat the gas.  We suggested that, from 
well-known gravitational dynamics, heating of the gas would slow down its contraction, and we used the words impede, prevent, and hinder the contraction to describe the 
effect on the gas. In Paper I we estimated, within a simple spherically-symmetric model, that for the case of a 100 GeV  
WIMP, annihilation heating would overtake cooling at gas densities of the order of $10^{13}~{\rm cm}^{-3}$. We considered both fully dissociated H$_2$ and 100\% of 
hydrogen in H$_2$, showing that in both cases heating from annihilation will eventually come to dominate over cooling, although at different times. In paper I, we 
speculated what may happen after the annihilation heating overtakes the cooling. We suggested a few possibilities, among them the possibility that the gas continues to 
contract although at a slower rate, and the possibility that objects in hydrostatic equilibrium would form supported by the annihilation energy of the contracted dark matter 
instead of nuclear fusion. It is the latter objects that we called dark stars.

In subsequent papers (Freese et al.\ 2008; Spolyar et al.\ 2009), we examined the theoretical existence of equilibrium stellar objects supported by dark matter annihilation only, and we studied their properties.  We found that the equations of hydrostatic and gravitational equilibrium admit dark star solutions with specific relations between their mass, radius, and energy input from annihilation. To fix the mass and the annihilation power, we focused on the dark matter and gas densities in the environment relevant for the first stars, as given by numerical simulations. 

Already in our second paper (Freese et al.\ 2008) we realized that, although dark matter heating beats cooling at a hydrogen density of $n_H=10^{13}$ cm$^{-3}$ and a core radius of 17 AU (for 100 GeV WIMPs), this density is too low to admit equilibrium dark star solutions.  The collapsing molecular cloud must contract further to $n_H \sim 10^{17}$ cm$^{-3}$ and a radius $\sim 2$ AU before an equilibrium dark star can come into existence.  
To find  an equilibrium DS, we considered as our starting point a (roughly) 3 $M_\odot$ polytropic object and varied the radius (iteratively) until both hydrostatic and thermal equilbrium were satisfied. We took the object with these properties as our initial model (defined to be $t=0$).  More specifically, for a 100 GeV WIMP with annihilation cross section $\langle \sigma v \rangle = 3 \times 10^{-26} ~{\rm cm^3/s}$, one of the first equilibrium dark stars we found had a mass of 3.96 $M_\odot$,  a radius of 1.95 AU, a surface temperature of 4,100 K, a central temperature of $6.3\times10^4$ K, a central gas density of $3.5\times10^{-7}$ g~cm$^{-3}$, a luminosity of $4.5 \times 10^4 ~L_\odot$, and a fully convective structure. For a 10 GeV WIMP mass, the values were 3.83 $M_\odot$, 3.5 AU, 4,100 K, $3.2\times10^4$ K, $4.8\times10^{-8}$ g/cm$^3$, $1.5\times10^5 ~ L_\odot$, respectively.  Converting to hydrogen number density, these numbers correspond to an average hydrogen density of $7.6 \times 10^{16}$ cm$^{-3}$ for the 100 GeV WIMP case ($1.2\times10^{16}$  cm$^{-3}$ for the 10 GeV WIMP case), and a central hydrogen density $n_H(0)= 3.6 \times 10^{17}$ cm$^{-3}$ for the 100 GeV case ($4.9 \times 10^{16}$ cm$^{-3}$ for the 10 GeV case).  In this conversion, we have used
$n_H(0) = \rho_B(0)/(\mu m_H)$, where $m_H$ is the proton mass, $\rho_B(0)$ is the central baryon density, 
and $\mu$ is the mean atomic weight.  Since H and He are fully ionized  (except near the surface) $\mu = 0.588$.  We note that the initial radius for the equilibrium dark star is $\sim$ 2--3 AU. 
 Hence the DS radius is much smaller than 17 AU, the size of the still collapsing core found in Paper I at the time DM heating first beats cooling.  
 Henceforth in this paper we will focus on the case of the 100 GeV WIMPs.  Though the detailed numbers will change for other WIMP masses, 
 our conclusions remain the same for masses ranging from 1 GeV-10 TeV.

In short, our earliest work on equilibrium dark stars illustrated that the initial hydrogen densities of the first dark stars that form are quite high, roughly $n_H \sim 10^{17}$ cm$^{-3}$ for the case of 100 GeV WIMPs.

The dark star solutions we found had surface temperatures much lower than the Pop III case without dark matter annihilation. As a consequence, dark stars produce a negligible amount of ionizing UV photons. In the absence of the latter,  gas could accrete onto the dark star and allow it to grow in time. We allowed mass to fall onto the star, one solar mass at a time, starting from our initial 3 $M_\odot$ dark star.  With each addition of mass, we allowed the radius to vary so as to look for new equilibrium solutions.  These have been published in tables in Spolyar et al.\ (2009). 
For example, we grow the DS to 100 M$_\odot$ where the central hydrogen density reaches $n_H=10^{18}$ cm$^{-3}$.  We continue to allow the DS to grow from there in a manner described in our other papers (Freese et al.\ 2008, Spolyar et al.\ 2009).  The DS continues to grow as long as it has DM fuel.  As a consequence it grows to at least $\sim 1000 ~M_\odot$, ten times the mass of Pop III stars in the standard picture (Freese et al.\ 2008, Spolyar et al.\ 2009, Sivertsson \& Gondolo 2011), and in some models even higher (Umeda et al.\ 2009, Freese et al.\ 2010).\footnote{If multiple protostars were present, Stacy et al.\ (2012) found that dynamical interaction might rapidly displace dark stars from the dark matter peak, thereby depriving them prematurely of the DM fuel. Smith et al.'s results do not support this scenario.} The ionizing photons from the much hotter standard Pop III stars provide feedback (Tan \& McKee 2004, McKee \& Tan 2008) that prevents them from growing  as large as in the case when the dark matter heating is included.

Subsequent authors have somehow missed the fact that equilibrium DS have the high densities they do.  These authors have run simulations up to $n_H=5 \times 10^{14}$ cm$^{-3}$ in the presence of annihilation  heating, and have seen no sign that the protostellar cloud stops collapsing.  We completely agree. These densities are simply too low.  Though heating does dominate at that point, the radius must contract further to allow for equilibrium.\footnote{In Paper I we had not yet constructed equilibrium dark stars and did not yet know these numbers.}  These authors argued that we were wrong about the formation of DS, but their arguments are incorrect.  Their simulations stopped many orders of magnitudes short of ever reaching the densities of DS.  For example, Ripamonti et al.\ (2010) stopped at $n_H=10^{14}$ cm$^{-3}$.  We would agree that DS have not formed yet at those densities:  the gas must contract further to $10^{17}$ cm$^{-3}$.  A related issue is the DS radius.  Whereas in Paper I we find that DM heating beats cooling inside a core of $\sim$ 20 AU, in fact the star must contract further to become a DS, down to $\sim 2$ AU.  Smith et al.\ confuse these two numbers in remarking that our papers claim DS formation at 20 AU.  Their simulations show that collapse continues to smaller radii than 20 AU and we completely agree. This continued collapse merely agrees with the DS forming later on scales smaller than the resolution limit of Smith et al.

Smith et al.\ have run simulations of the evolution of the contracting gas under the influence of heating from dark matter annihilation.  Their very impressive simulations include a fully time-dependent chemical network, H$_2$ line cooling and collision-induced emission, cooling due to the collisional ionization and recombination of H and He, and heating and cooling due to changes in the molecular fraction, shock heating, compressional heating, and cooling due to rarefactions. As their initial conditions, they start from central regions (2 parsecs in size) of mini haloes with central hydrogen density $n_H = 10^6$ cm$^{-3}$ obtained from the simulations of Greif et al.\ (2011). Then they follow the contraction of this gas and show results at a hydrogen density of $\sim 5\times10^{14}~{\rm cm}^{-3}$. This  density is slightly higher than the value at which annihilation heating prevails over cooling (as shown in our original estimates and confirmed by Smith et al.).  However, they are apparently unable to reach the $\sim 10^{17}~{\rm cm}^{-3}$ density of our hydrostatic dark star solutions.   Hence, as stated above, we are in complete agreement with their physical result that the gas keeps collapsing to higher densities than the value of $n_H=3 \times 10^{14}$ cm$^{-3}$, which is past the point where dark matter heating dominates.   However we are in disagreement with some of the words in their paper which imply that dark stars are not formed.
The gas continues to collapse to several orders of magnitude higher than the limits of their simulations, at which point dark stars in thermal and hydrostatic equilibrium form. 

Although Smith et al.'s SPH simulations are state-of-the-art, they do not have the resolution to follow the detailed formation of a star, and they use  the sink particle approach to treat the gas that collapses to scales smaller than their resolution limit. In this approach (Bate et al.\ 1995), a region of the highest density in the 3D simulation --- a protostellar cloud --- is replaced by a single non-gaseous particle which contains all the mass in the region and accretes any infalling mass.  Sink particles accrete gas particles that come within a predetermined accretion radius provided they satisfy appropriate criteria.  The simulation continues as usual for the rest of the gas, but with the sink particle replacing the highest density gas regions. This allows the gas evolution outside the sinks to be followed beyond the time of sink formation. Apart from their masses, the characteristics of the sinks have to be set by hand and do affect the gas around them.   Smith et al.\ take the sinks to have an accretion radius of $\sim 5$ AU and treat the dark matter as glued to the first sink that forms. They further make the assumption that the sinks can be treated as ordinary fusion-powered Pop III stars.  They then stop their simulations at a time just before such Pop III stars would emit ionizing UV photons that would prevent further accretion (Tan \& McKee 2004, McKee \& Tan 2008). At this point the sinks have accreted masses of roughly $10 ~M_\odot$.   However, we believe that these sinks instead contain dark stars (or, depending on their density, are themselves dark stars).  The sinks are dense enough and large enough that dark matter annihilation products would be trapped inside them and would heat the gas inside the sink.  Inside the sinks the gas could collapse a bit further until it became a dark star at $n_H \sim 10^{17}$ cm$^{-3}$.  The gas would not contract far enough to ignite fusion.  Thus the sinks should be treated as dark stars, which are much cooler, never produce ionizing photons, and can grow much larger, up to at least $\sim$ 1000 M$_\odot$ or more.  

Smith et al.\ find the interesting result that a $\sim 1000$ AU gas disk  forms around the central object due to the dark matter heating. Smith et al.\ observe that at the outskirts of the disk, ionization from dark matter annihilation leads to an increase in the production of H$_2$. The extra H$_2$ cools the gas around the disk, makes it fall onto the disk faster, and makes it pile up against the slower gas in the disk.  Previous authors (Clark et al.\ 2008, 2011a, 2011b; Turk et al.\ 2009; Stacy et al.\ 2010; Greif et al.\ 2011, 2012) argued that the protostellar accretion disk that builds up around the first protostar in the absence of dark matter heating rapidly becomes unstable and fragments.  However, Smith et al.\ show that, with the inclusion of dark matter heating, the disk is stabilized against fragmentation.  Thanks to the annihilation heating, the disk has Toomre parameter $Q>1$, which makes it a ``stable disk,'' in the sense that fragmentation is suppressed.  Smith et al.\ find that additional sinks (in addition to the original sink at the center of the mini halo) form only far from the center, at a distance of 1000 AU from the primary.  Compared to the case without dark matter heating, only a few sinks form,  far apart from each other, leading Smith et al.\ to their main conclusion: heating from dark matter annihilation suppresses fragmentation in primordial star formation. 

It seems to us that Smith et al.'s work supports the dark star scenario.  The gas collapses down to a radius where an equilibrium dark star can form.  This dark star is effectively their sink.  Then an accretion disk forms as more mass falls onto the dark star. In their words, the sink mass grows.  Whereas one might worry that fragmentation might perturb the dark matter away from the dark star and remove its power source (Stacy et al.\ 2012), instead Smith et al.\ show that this does not occur. Instead, any further sinks form only far enough away as to leave the dark star undisturbed in the comfort of its dark matter surroundings.  

In short, the simulations of Smith et al.\ not only confirm our predictions about dark stars in the range where the simulations apply, but also solve a potential obstruction to dark star formation by showing that dark matter annihilation prevents the fragmentation of the collapsing gas.
 
To conclude this paper, we recommend some future improvements required to fully understand the formation and evolution of dark stars.  First, higher resolution simulations 
are needed to study the details of the contraction of the gas, including its angular momentum, and including the numerical evolution of a dynamical dark matter halo, all the way to the formation of equilibrium dark stars.  At present this regime is an interesting unstudied missing piece of the DS formation. Second, we suggest as an important improvement to the work of Smith et al., a better treatment of the energy deposition in the collapsing gas phase (prior to the formation of DS).  Whereas these authors treated the energy deposition using a split between ionization and heating appropriate to energies below 10 keV or so (Vald\'es \& Ferrara 1995), at the higher energies of DM annihilation products the situation is more complicated.   WIMPs of 100 GeV annihilate to particles with energies far above 1 GeV, which behave very differently.    For example, photons and electron/positron pairs above 100 MeV produce electromagnetic cascades in which photons pair produce $e^+/e^-$ pairs; these produce bremstrahlung photons, etc. The details of the high energy annihilation products of dark matter should be simulated.

\end{document}